\documentclass[twocolumn,amsmath,amssymb]{revtex4}

\usepackage{graphicx}
\usepackage{dcolumn}
\usepackage{bm}
\newcommand{\etal}{\emph{et al.}}
\newcommand{\be}{\begin{equation}}
\newcommand{\ee}{\end{equation}}
\newcommand{\bfig}{\begin{figure}}
\newcommand{\efig}{\end{figure}}
\newcommand{\incl}{\includegraphics}

\begin{document}      

\title{Low temperature vortex liquid in $\rm La_{2-x}Sr_xCuO_4$
} 
\author{Lu Li$^1$, J. G. Checkelsky$^1$, Seiki Komiya$^2$, 
Yoichi Ando$^2$, and N. P. Ong$^1$\footnote{preprint submitted to Nature Physics}
}
\affiliation{
$^1$Department of Physics, Princeton University, Princeton, NJ 08544, USA\\
$^2$Central Research Institute of Electric Power Industry, Komae, Tokyo 201-8511, Japan
} 

\date{\today}     
\begin{abstract}
\end{abstract}

\maketitle                   
In the cuprates, the lightly-doped region is of major
interest because superconductivity, antiferromagnetism,
and the pseudogap state~\cite{Timusk,Lee,Anderson} come together near a critical
doping value $x_c$. These states are deeply influenced by
phase fluctuations~\cite{Emery} which lead to a vortex-liquid state
that surrounds the superconducting region~\cite{WangPRB01,WangPRB06}. 
However, many questions~\cite{Doniach,Fisher,FisherLee,Tesanovic,Sachdev} related to the nature of the
transition and vortex-liquid state at very low tempera-
tures $T$ remain open because the diamagnetic signal is
difficult to resolve in this region. Here, we report torque
magnetometry results on $\rm La_{2-x}Sr_xCuO_4$ (LSCO) which
show that superconductivity is lost at $x_c$ by quantum
phase fluctuations. We find that, in a magnetic field $H$,
the vortex solid-to-liquid transition occurs at field $H_m$
much lower than the depairing field $H_{c2}$. The vortex
liquid exists in the large field interval $H_m \ll H_{c2}$, even in
the limit $T\rightarrow$0. The resulting phase diagram reveals the
large fraction of the $x$-$H$ plane occupied by the quantum
vortex liquid.  

In underdoped LSCO, the magnetic susceptibility is
dominated by the Curie-like spin susceptibility and the
van Vleck orbital susceptibility~\cite{Keimer,Niedemayer}. These large para-
magnetic contributions render weak diamagnetic signals
extremely difficult to detect using standard magnetome-
try in lightly-doped crystals. However, because the spin
susceptibility is nearly isotropic~\cite{Ando} while the incipient
diamagnetism is highly anisotropic (the supercurrents are
in-plane), torque magnetometry has proved to be effective in resolving 
the diamagnetic signal~\cite{Farrell,Bergemann,WangPRL05,Li05}. 
With $\bf H$ tilted at a slight angle $\phi$ to the crystal $c$-axis, the torque $\tau$ may be
expressed as an effective magnetization
$M_{obs}\equiv\tau/\mu_0H_xV$, where $V$ is the sample volume, $\mu_0$ the permeability 
and $H_x = H\sin\phi$ (we take $\bf \hat{z}\;||\;\hat{c}$).  In cuprates,
$M_{obs}$ is comprised of 3 terms~\cite{WangPRL05,Li05}
\be
M_{obs}(T,H_z) = M_d(T,H_z) + \Delta M_s(T,H_z) + \Delta\chi^{orb}(T)H_z,
\label{M3}
\ee
where $M_d(T,H_z)$ the diamagnetic magnetization of interest,
$\Delta M_s$ the anisotropy of the spin local moments, and 
$\Delta\chi^{orb}$ the anisotropy of the van Vleck 
orbital susceptibility [see SI].  Hereafter, we write $H$ for $H_z$.

We label the 7 samples studied as 03 (with x = 0.030), 04 (0.040),
05 (0.050), 055 (0.055) 06 (0.060), 07 (0.070) and 09
(0.090). To start, we confirmed that, above ~$\sim$25 K,
$M_{obs}$ derived from the torque experiment in sample 03
is in good, quantitative agreement with the anisotropy
inferred from previous bulk susceptibility measurements
on a large crystal of LSCO (x = 0.03)~\cite{Ando} (see SI for
comparisons).

Figure \ref{Mobs} displays the magnetization $M_{obs}$ in samples 055 and 06.   
The pattern of $M_{obs}$ results from the sum of the 3 terms in Eq. \ref{M3}. 
Panel (a) shows how it evolves in sample 055.  At high $T$
(60--200 K), the curves of $M_{obs}$ vs. $H$ are fan-like, 
reflecting the weak $T$ dependence of the orbital term $\Delta\chi^{orb}(T)H$~\cite{WangPRL05}.
At the onset temperature for diamagnetism $T_{onset}$ (55 K, bold curve), the diamagnetic
term $M_d$ appears as a new contribution.  The strong $H$ dependence of $M_d$ 
causes $M_{obs}$ to deviate from the $H$-linear behavior.
In Panel b, the evolution is similar, except that the larger diamagnetism
forces $M_{obs}$ to negative values at low $H$.  As mentioned, the spin contribution $\Delta M_s$
is unresolved above $\sim$40 K in both panels.  To magnify the diamagnetic signal, 
it is convenient to subtract the orbital term $\Delta \chi_{orb}H$.

The resulting curves $M_{obs}'(T,H) \equiv M_{obs}-\Delta\chi^{orb}H$ are shown 
for sample 05 in Panel (c).
At low fields $M_{obs}'$ displays an interesting oscillatory behavior (curves at 0.5 and 0.75 K),
but at high fields it tends towards saturation.  By examining how $M_{obs}'$ behaves in the 2 limits of weak and 
intense fields in the 7 samples (see SI), we have found that $M_{obs}'$ is comprised of 
a diamagnetic term $M_d(T,H)$, that closely resembles the ``tilted hill'' profile of diamagnetism
in the vortex liquid state above the critical temperature $T_c$ reported previously~\cite{WangPRL05,WangPRB06}, 
and a spin-anisotropy term $\Delta M_s$ that becomes large at low $T$.  Modeling 
the latter as free spin-$\frac12$ local moments with anisotropic $g$
factors measured with $\bf H || c$ ($g_c$) and $\bf H\perp c$ ($g_{ab}$), we have
(details in SI)
\be
\Delta M_s(T,H) = {\cal P}(T)\tanh[\beta g_{\phi}\mu_BB/2],  
\label{Ms}
\ee
with $\mu_B$ the Bohr magneton, $\beta = 1/k_BT$ 
and $g_{\phi} = \sqrt{(g_c\cos\phi)^2+(g_{ab}\sin\phi)^2}$.
With $g_{\phi}\sim g_c$ fixed at 2.1, the sole adjustable parameter at each $T$ is 
the prefactor ${\cal P}(T)$.  

Equation \ref{Ms} 
accounts very well for the curves in Fig. \ref{Mobs}c, especially the oscillatory behavior
and the saturation at large $H$: at $T$ = 0.5 and 0.75 K, $\Delta M_s\sim 1/k_BT$ dominates $M_d$ 
in weak $H$, but for $H>k_BT/g_c\mu_B$, the
saturation of $\Delta M_s$ implies that $M_{obs}'(H)$ adopts the profile of $M_d(H)$ apart from 
a vertical shift.  Lightly doped LSCO enters a spin- or cluster-glass state~\cite{Keimer,Niedemayer} 
below the spin-glass temperature $T_{sg}$ which is sensitive to sample purity 
(in our crystals 03 and 04, $T_{sg}\sim$ 2.5 and 1 K, respectively).  The
magnetic hysteresis below $T_{sg}$ (clockwise) is distinct from 
the hystereses (anticlockwise) in the vortex solid, and is significant in only these 2 samples.

\begin{figure*}		
\incl[width=18.5cm]{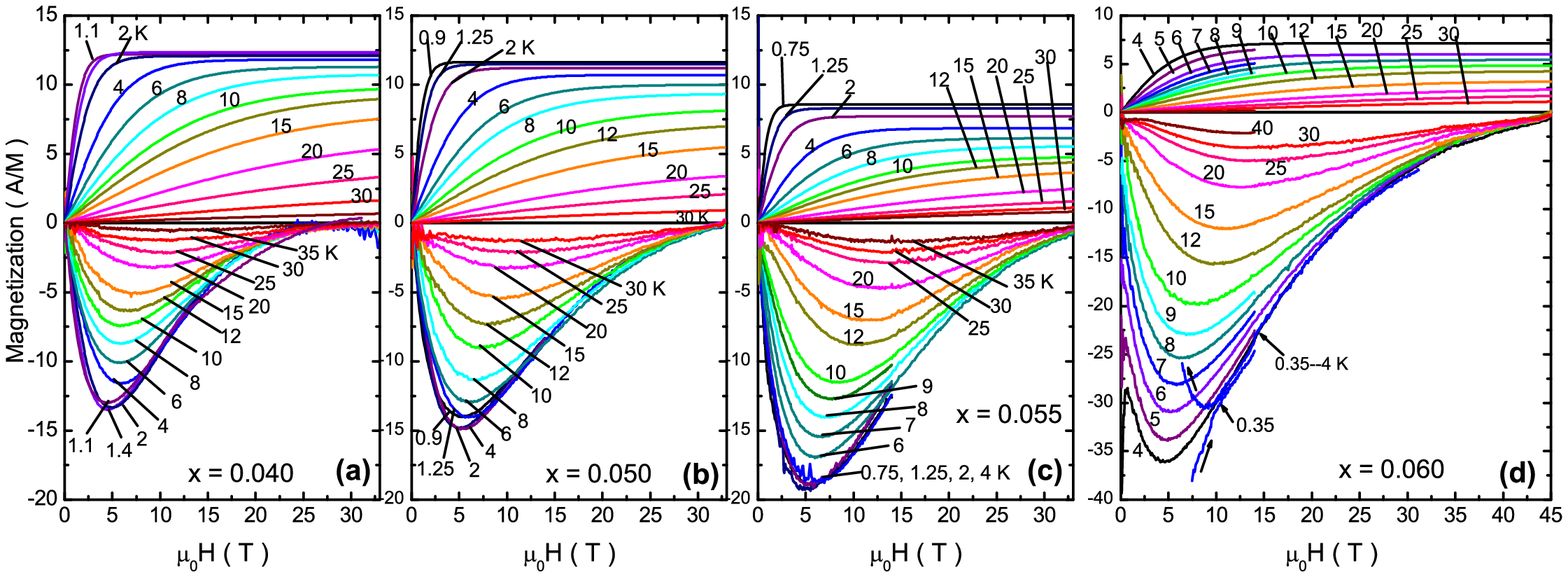}   
\caption{\label{Md} 
The paramagnetic spin term $\Delta M_s(T,H)$ and the diamagnetic 
term $M_d(T,H)$ vs. $H$ in samples 04, 05, 055 and 06
(Panels a--d, respectively).  In each panel, the diamagnetic minimum (at 5 T) 
deepens rapidly between 30 K and 5 K, but ceases to change below $T_Q$.
The depairing field $H_{c2}$ is estimated to be 25, 35, 43, and 48 T in Panels (a)--(d), respectively. 
In Panel d, the branching curves (with arrows) indicate the high-field limit of the vortex solid 
at 0.35 K.  Above $H_{irr}(T)$, the low-$T$ curves merge with the vortex-liquid curve at 4 K.
}
\end{figure*}

Subtracting $\Delta M_s$ from $M_{obs}'$, we isolate the purely diamagnetic
term $M_d(T,H)$.   In Fig. \ref{Md}, we display the curves of $M_d$ and $\Delta M_s$
at selected $T$ in samples 04, 05, 055 and 06.  The samples 03,
04 and 05 do not display any Meissner effect at all.  The strict reversibility of 
the $M_d$--$H$ curves confirms that we are in the vortex-liquid state in 03, 04 and 05. 
When $x$ exceeds $x_c$,
the samples display broad Meissner transitions ($T_c\sim$ 0.5 and 5 K 
in 055 and 06, respectively).  Hysteretic behavior appears below a strongly
$T$-dependent irreversibility field $H_{irr}(T)$, which we discuss shortly.  
Examination of the $T$ dependences in the 4 panels uncovers an important pattern.
In the vortex liquid, the overall magnitude of $M_d$ grows rapidly as we cool from 35 to 5 K, but
it stops changing below a crossover temperature $T_Q$ ($\sim$ 4 K in samples 05, 055 and 
06, and $\sim$2 K in 04).  Even in 06, where $H_{irr}\sim$ 9 T at 0.35 K,
$M_d$ recovers the $T$-independent profile when $H>H_{irr}$ (note the diverging
branches at 9 T in Panel d).  The insensitivity to $T$ suggests that the excitations, 
which degrade the diamagnetic response in the liquid state, are governed by quantum 
statistics below $T_Q$.  

In intense fields, the field suppression of $M_d$ provides
an estimate of the depairing field $H_{c2}$ ($\sim$20, 25, 35, 43, and 48 T
in samples 03, 04, 05, 055, and 06, respectively).  We find that $H_{c2}$ is 
nominally $T$-independent as reported earlier~\cite{WangPRL05}.  

Experimentally, the appearance of hysteresis in $M_d$ vs. $H$ 
below $H_{irr}(T)$ is a sensitive barometer of the vortex
solid. The strong vortex pinning in LSCO leads to large
hystereses as soon as the vortex system exhibits shear
rigidity. The hysteretic loops, which appear in 055 ex-
pand very rapidly as $x$ exceeds 0.055. By plotting the
hysteretic loops in magnified scale (Fig. \ref{Hirr}a shows curves
for 06), we can determine $H_{irr}(T)$ quite accurately. Vor-
tex avalanches -- signatures of the vortex solid -- are observed 
(for $H<H_{irr}$) unless the field-sweep rate is very
slow (see SI).

The temperature dependence of $H_{irr}(T)$ is plotted in Fig. \ref{Hirr}b for samples $x>x_c$.
At low $T$, the dependence approaches the exponential form 
\be
H_{irr}(T) = H_0\exp(-T/T_0).  
\label{DW}
\ee
The parameters $H_0$ and $T_0$ decrease steeply as $x\rightarrow x_c$.  
Previous experiments in cuprates were not 
performed to low enough $T$ or to high enough $H$ to observe the exponential form.  
The field parameter $H_0$ provides an upper bound for the zero-Kelvin melting field $H_m(0)$
(at some $T<$0.35 K, crossover to a quantum melting may
cause $H_{irr}$ to deviate from Eq. \ref{DW},
so $H_0$ here is a close upper bound to $H_m(0)$).  Equation \ref{DW} is
reminiscent of the Debye-Waller factor, and strongly suggests that the excitations responsible 
for the melting transition follow classical statistics at temperatures down to 0.35 K.  The 
classical nature of these excitations contrasts with the 
quantum nature of the excitations in the vortex liquid below $T_Q$ described above.

The inferred values of $H_0$ (2, 13, 25 and 40 T in 055, 06,
07, and 09, respectively) are much smaller than $H_{c2}(0)$.
Hence, after the vortex solid melts, there exists a broad
field range in which the vortices remain in the liquid state
at low $T$. The existence of the liquid at $T<T_Q$ implies
very large zero-point motion associated with a small vortex 
mass $m_v$, which favors a quantum-mechanical description. 

Finally, we construct the low-$T$ phase diagram in the
$x$--$H$ plane. Figure \ref{phase} shows that the $x$ dependence of
$H_{c2}(0)$, the depairing field scale, is qualitatively distinct
from that of $H_0$, the boundary of the vortex solid. The
former varies roughly linearly with $x$ between 0.03 and
0.07 with no discernible break-in-slope at $x_c$, whereas
$H_0$ falls steeply towards zero at $x_c$ with large negative
curvature. This sharp decrease -- also reflected in the
1000-fold shrinkage of the hysteresis amplitude between
$x$ = 0.07 and 0.055 -- is strong evidence that the collapse
of the vortex solid is a quantum critical transition. This
is shown by examining the variation of $H_{irr}$ vs. $x$ at several 
fixed $T$ (dashed lines). At 4 K, $H_{irr}$ approaches 0
gently with positive curvature, but at lower $T$, the trajectories 
tend towards negative curvature. In the limit
$T$ = 0, $H_0$ approaches zero at $x_c$ with nearly vertical
slope. The focussing of the trajectories to the point ($x_c$, 0) 
is characteristic of a sharp transition at $x_c$, and strikingly 
different from the smooth decay suggested by viewing 
lightly-doped LSCO as a system of superconducting
islands with a broad distribution of $T_c$'s.

In Fig. \ref{phase} the high-field vortex liquid is seen to extend 
continuously to $x<x_c$ where it co-exists with the
cluster/spin-glass state~\cite{Keimer,Niedemayer} (samples 03, 04 and 05).
As shown in Fig. 2 (see SI), the robustness of $M_d$
to intense fields attests to unusually large pairing energy
even at $x$ = 0.03, but the system stays as a vortex liquid
down to 0.35 K.

In the limit $H\rightarrow$0, the vortex liquid ($x<x_c$) has
equal populations of vortices and antivortices. This implies that, if $x$ 
is reduced below $x_c$ at low $T$ and in zero
field, superconductivity is destroyed by the spontaneous
appearance of free vortices and antivortices engendered
by increased charge localization and strong phase fluctuation~\cite{Doniach,Fisher,Tesanovic}. 
The experiment lends support to the
picture that, at $T$ = 0 in zero field, superconductivity
first transforms to a vortex-liquid state that has strong
pairing but lacks phase coherence before the insulating
state is attained. The rapid growth of the spin/cluster-
glass state in LSCO suggests that incipient magnetism
also plays a role in destroying superconductivity.
In summary, we find that the pair condensate in LSCO
is robust even for $x$ = 0.03 in fields of 25 T and higher.
However, because the vortex solid melts at a lower field
$H_0$, the condensate exists as a vortex liquid that resists
solidification down to 0.35 K, implying large zero-point
motion consistent with a small vortex mass. In the phase
diagram (Fig. \ref{phase}), the vortex liquid surrounds the vortex
solid region. The evidence supports a sharp quantum
critical transition at $x_c$. However, the $T$ dependence in
Eq. \ref{DW} implies that the quantum melting of the solid must
occur below 0.35 K even as $x\rightarrow x_c$.


{\bf Acknowledgements}
Valuable discussions with Yayu Wang, Z. Te\v{s}anovi\'c, S. Sachdev, 
S. A. Kivelson, P. W. Anderson and J. C. Davis are acknowledged.  The research at Princeton 
was supported by the National Science Foundation (NSF) through a MRSEC grant DMR 0213706.
Research at CRIEPI was supported by a Grant-in-Aid for Science from the Japan Society for 
the Promotion of Science.  The high field measurements were performed in the 
National High Magnetic Field Lab. Tallahassee, which is supported by NSF, the 
Department of Energy and the State of Florida.

\newpage
\bfig
\incl[width=9.5cm]{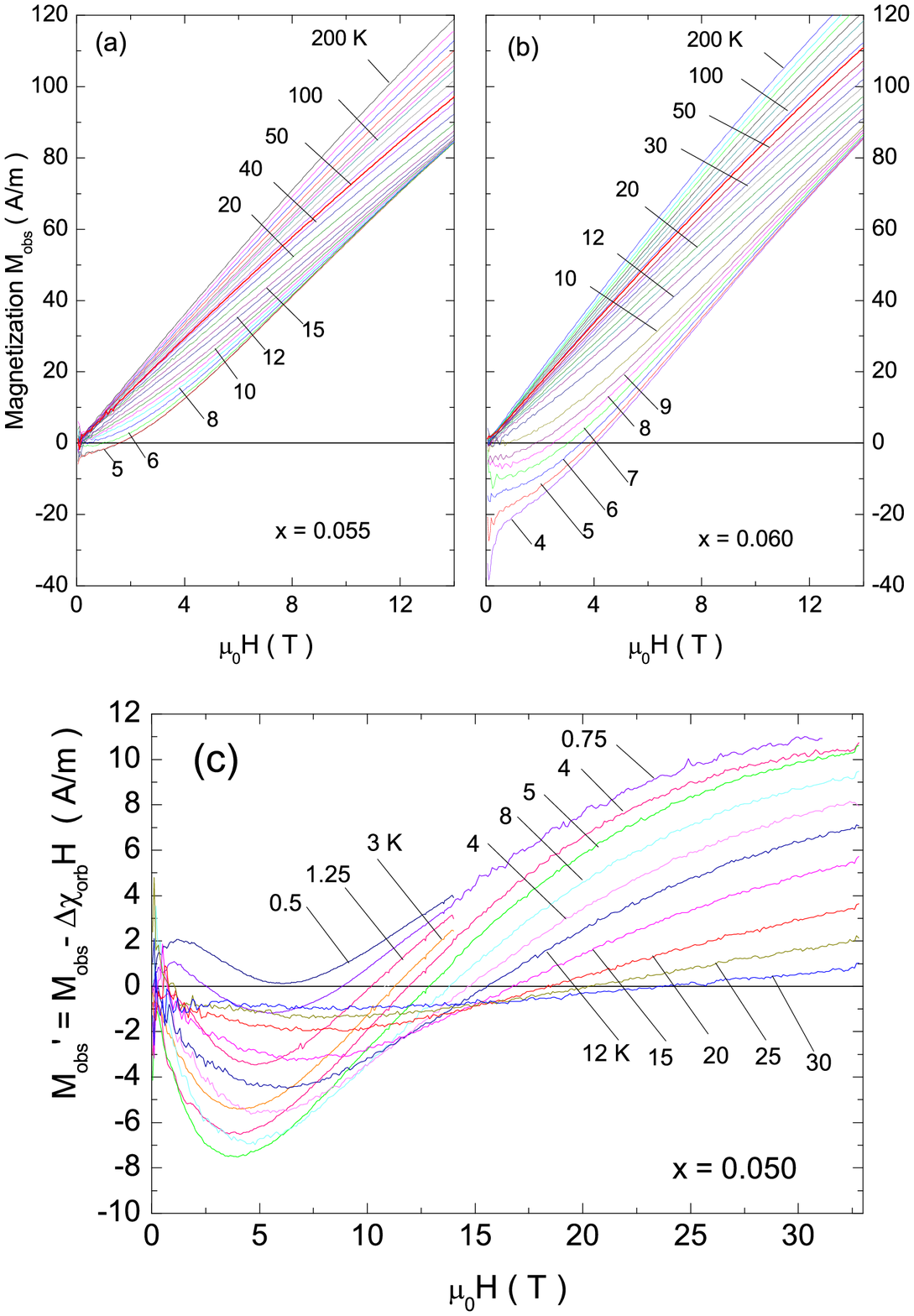}  
\caption{\label{Mobs} 
Curves of the observed magnetization $M_{obs}$ vs. $H$ at temperatures
4--200 K in sample 055 (Panel a, $T_c\sim$ 0.5 K) and 06 (Panel b, $T_c\sim$ 5 K).  
The crystal is glued to the tip of the cantilever with its $c$-axis at a small 
angle $\phi\sim 15^{\bf o}$ to the field $\bf H$.
Above $T_{onset}$ (bold curves at 55 and 70 K in a and b, respectively), the fan-like
pattern is due entirely to the paramagnetic term $\Delta\chi^{orb}(T)H$ (see SI
for plot of $\Delta\chi(T)$).  Below $T_{onset}$, the diagmagnetic 
term $M_d$ becomes evident.  Panel (c) shows the profiles of 
$M_{obs}'= M_{obs}-\Delta\chi^{orb}H$ in the sample 05.  Note the oscillation in 
weak $H$ at $T$ = 0.5 and 0.75 K and the approach towards saturation in high fields.  
These curves are separated into $\Delta M_s$ and $M_d$ in Fig. \ref{Md}b.
}
\efig
\bfig			
\incl[width=9cm]{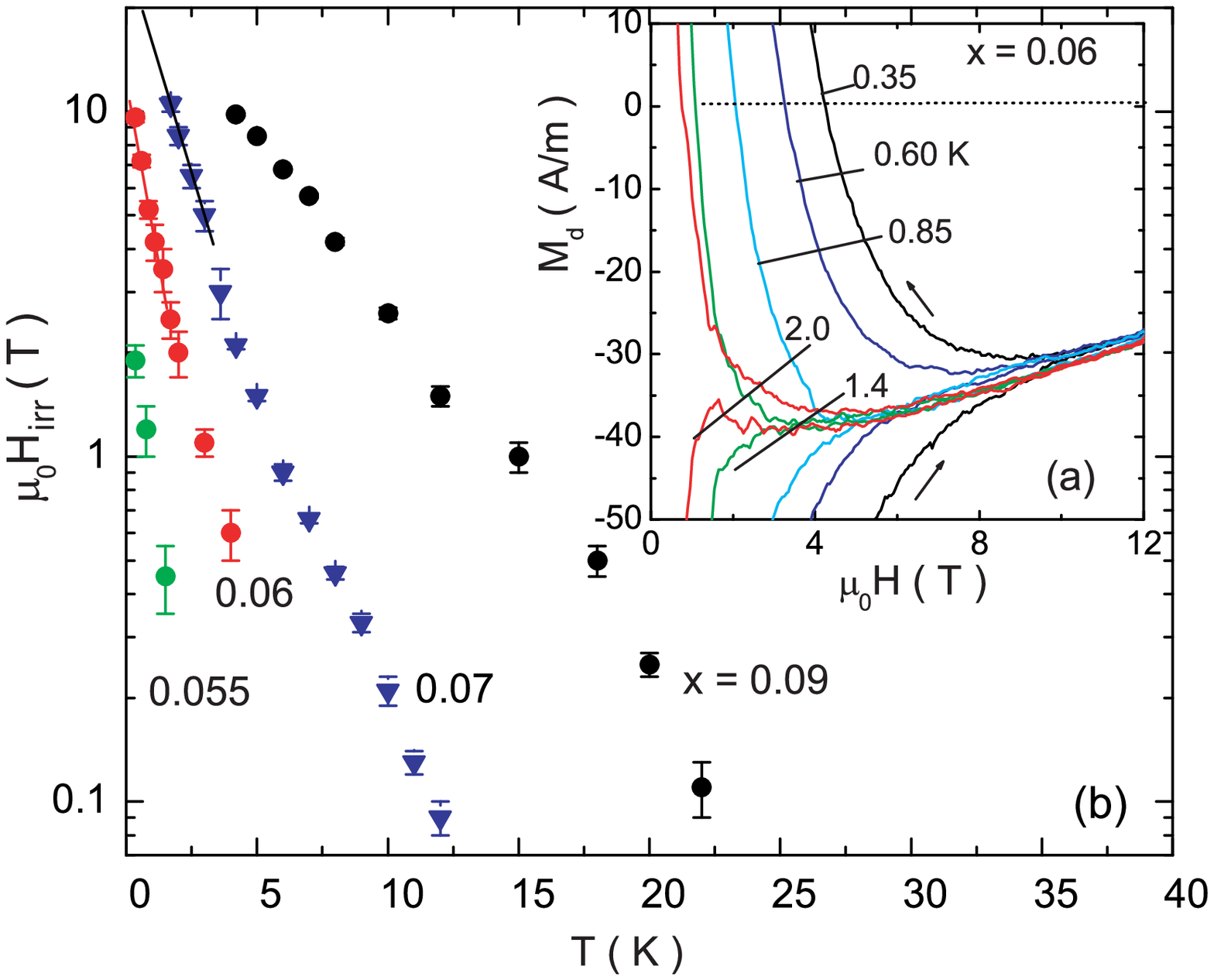}   
\caption{\label{Hirr} 
The hysteretic curves in the vortex-solid phase of LSCO (Panel a) and 
the $T$ dependence of the irreversiblity field $H_{irr}(T)$ in several samples (Panel b)
Panel a displays hysteretic curves in Sample 06 at $T$ from 0.35 to 2 K.  
Although the hysteretic segments for $H<H_{irr}(T)$ are very strongly $T$
dependent, the reversible segments above $H_{irr}(T)$ are not.  The latter match the $T$-constant profile 
shown in Fig. \ref{Md}d.  The $T$ dependences of $H_{irr}(T)$ in the samples 055, 06, 07 and 09
are shown in semilog scale in Panel b.  At low $T$, the data approach
Eq. \ref{DW}.  The steep decrease of the characteristic temperature $T_0$ as $x\rightarrow x_c$
implies a softening of the vortex solid ($T_0\sim$1 K in 06).
}
\efig

\bfig	
\incl[width=9cm]{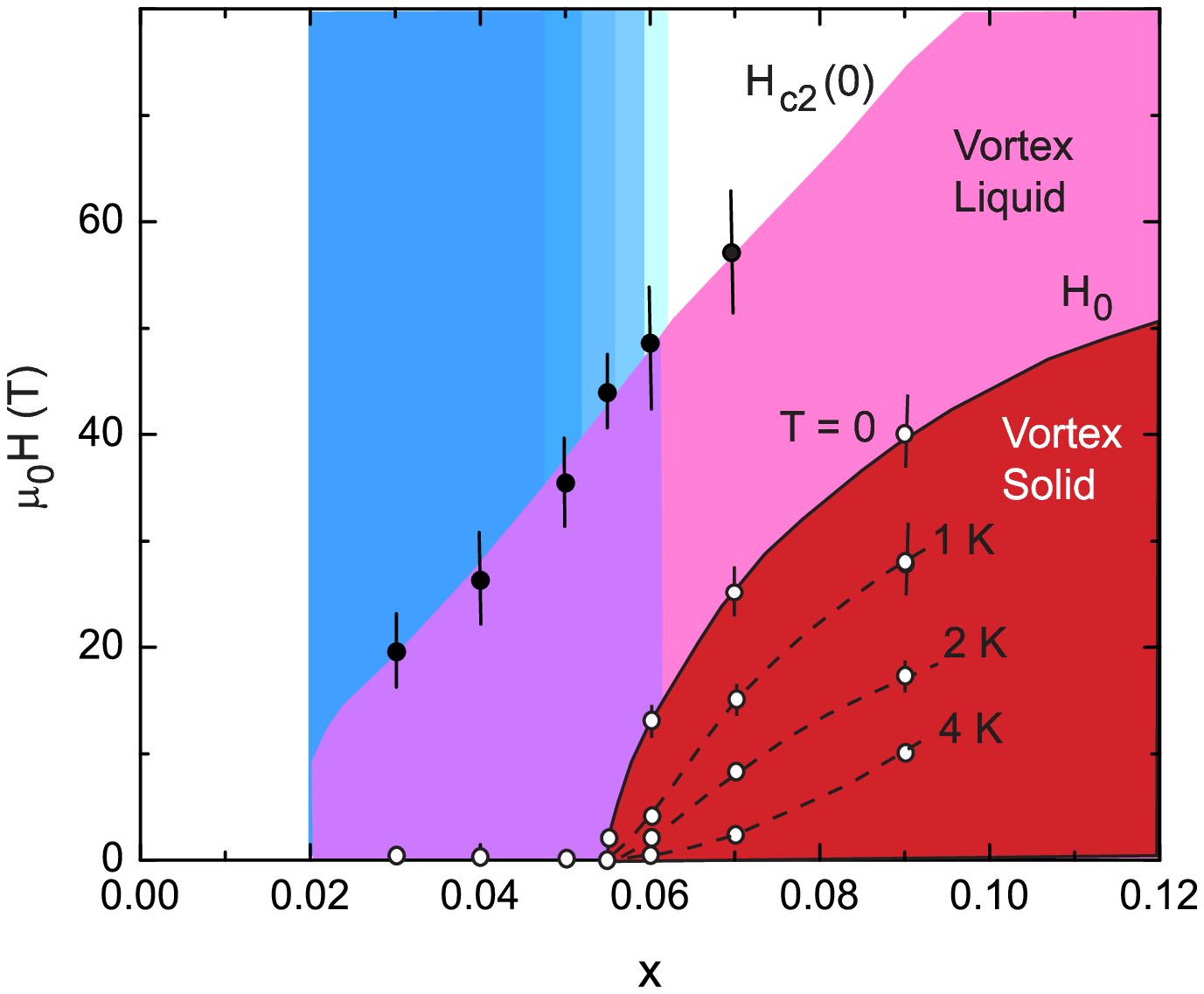}		
\caption{\label{phase}
The $x$--$H$ phase diagram of LSCO at low temperature, showing
the vortex-solid state and the vortex-liquid state.  The field $H_0 = \lim_{T\rightarrow 0} H_{irr}$
falls steeply to zero as $x\rightarrow x_c$(solid curve).  
The dashed lines indicate the variation
of $H_{irr}(T)$ vs. $x$ at fixed $T$, as indicated. By contrast, the
depairing field $H_{c2}(0)$ (closed circles) is nominally linear in
$x$. Below $x_c$, the vortex liquid is stable and coexists with a
growing magnetic background (graded shading).
}
\efig

\end{document}